\def\year{2017}\relax
\documentclass[letterpaper]{article}
\usepackage{aaai17}
\usepackage{times}
\usepackage{helvet}
\usepackage{graphicx}
\usepackage{courier}
\usepackage{booktabs} 
\usepackage{amsmath}
\usepackage{flexisym}
\usepackage{comment}
\usepackage{url}

\graphicspath{ {graphics/} }
\begin{document}
%
\title{Event Organization 101: Understanding
Latent Factors of Event Popularity}

\author{Shuo Zhang, Qin Lv\\
Computer Science Department\\
University of Colorado Boulder, Boulder, CO 80309 USA\\
Email: \{jasonzhang, qin.lv\}@colorado.edu\\
}

\maketitle

\begin{abstract}
The problem of understanding people's participation in real-world events 
has been a subject of active research and can offer valuable insights for 
human behavior analysis and event-related recommendation/advertisement. 
In this work, we study the latent factors for determining event popularity 
using large-scale datasets collected from the popular Meetup.com EBSN 
in three major cities around the world. We have conducted modeling 
analysis of four contextual factors (spatial, group, temporal, and 
semantic), and also developed a group-based social influence 
propagation network to model group-specific influences on events. 
By combining the Contextual features And Social Influence NetwOrk, 
our integrated prediction framework CASINO can capture the diverse 
influential factors of event participation and can be used by event 
organizers to predict/improve the popularity of their events. 
Evaluations demonstrate that our CASINO framework achieves high 
prediction accuracy with contributions from all the latent features 
we capture. 
\end{abstract}

\section{Introduction}
\label{sec: intro}

With the proliferation of event-based social networks (EBSNs) 
such as Meetup.com, Plancast.com, 
Douban Location (e.g., beijing.douban.com), 
and Facebook Events (events.fb. com), 
organizing and joining social events 
have become much easier than ever before. 
There are three key elements in the popular 
Meetup EBSN. 
{\em Users} can join different Meetup {\em groups}, which 
belong to different group categories and 
usually have specific themes such as hiking, writing, or health. Each 
group can organize various types of real-world {\em events} and encourage its 
group members to attend. 

Previous research has studied users' mobility or event participation 
behaviors in order to make personalized predictions or recommendations
~\cite{georgiev2014call,du2014predicting,macedo2015context}. 
For example, the work by Du et al. discovered a set of factors that will 
influence individual's attendance of activities, but the events they considered 
are organized by individuals, not groups~\cite{du2014predicting}.
Although those works shed some light on event organization, they focused  
on personalized prediction or recommendation by discovering individual 
users' preference profiles. To the best of our knowledge, no prior work has addressed the problem of 
identifying and combining the latent factors of group-organized event 
popularity to predict or improve the success of events organized by 
diverse social groups.

In this work, using two years of Meetup data collected in three major cities, 
we aim to capture the key factors that may impact the popularity of specific 
events organized by diverse social groups. 
By identifying and modeling the contextual factors along with group-based   
social influence on event participation, we propose an integrated framework 
CASINO to predict the popularity of group-organized events. Evaluations 
using large-scale Meetup data in three different cities demonstrate high 
accuracy of our method. We also compare the predictive power of the 
individual factors for different types of groups, which offer 
valuable insights for event organizers. 

\section{Data Collection and \\Problem Formulation}
\label{sec:data}

\subsection{Meetup Data Collection}
\label{sec: dataset}

Most historical 
information on Meetup is available via Meetup's streaming API~\footnote{https://www.meetup.com/meetup\_api/}. Using this 
API, we have collected comprehensive Meetup data from three cities: 
New York (NYC), London (LON) and Sydney (SYD) 
for the period of 
July 2013 to June 2015. Groups with less than 
15 events during the two-year period are considered inactive and 
removed during preprocessing. 
Table~\ref{tab: dataset} summarizes the key statistics of the three datasets. 

\begin{table}
	\centering
	\caption{Statistics of Meetup Datasets}
	\begin{tabular}{|c||rrrr|} \hline 
		City & \#groups & \#users & \#events & \#rsvps\\ 
		\hline \hline 
		New York & 2,802 & 248,211 & 270,321 & 1,613,634 \\ 
		\hline
		London & 1,534 & 155,883 & 117,862 & 945,669\\
		\hline
		Sydney & 706 & 55,768 & 55,295 & 353,149\\
		\hline
	\end{tabular}
\label{tab: dataset}

\end{table}

\subsection{Problem Formulation}
\label{sec: formulation}
Given a new event $e$ with its organizer, 
venue location, start time, title, description, and the group it belongs to, 
instead of predicting the absolute popularity in all events, 
our goal is to predict the relative popularity of events in the group category $c \in C$ 
that they belong to. The reason is that Meetup event sizes vary significantly 
across different group categories.  
We normalize event size by group category.  
Let $N_e$ be the number of attendees of an event $e$ and $avg_c$ be 
the average number of event attendees in group category $c$ that $e$ 
belongs to, we would predict the relative event popularity: $P_e = N_e/avg_c$.
In other words, we estimate the level of popularity of each event relative to 
other events in the same group category,  which is more informative and 
can offer more valuable insights for event organizers. 

\section{Contextual Features}
\label{sec: analysis}

In this section, we describe in detail our modeling analysis in order to 
understand and model the latent factors that can impact event popularity. 
Specifically, considering a group organizer who is planning a new 
event, we could potentially leverage the following 
information: spatial, group, temporal, and semantic features. 

\subsection{Spatial Features}

Choosing the right venue for an event is of particular importance in event 
organization. Intuitively, the event venue should be convenient for 
interested users (i.e., group members), yet not competing with too many 
other group events with similar themes. To model these influences, we 
propose the measures of location quality 
and competitiveness for each offline event.

\subsubsection{Location Quality}

Jensen's location quality has been widely used in analyzing static retail 
stores'  spatial interactions among different place 
categories~\cite{ICWSM148071,jensen2009analyzing}. 
We extend this method to 
our EBSN setting. We hypothesize that group categories will have similar 
attractiveness value between each other. We extend Jensen's inter 
coefficient to compute the relative number of events in other group categories 
that are near a given event. The value will be normalized compare with the 
scenario of placing all event locations uniformly random in the whole city area. 
Specifically, we first define the neighborhood event set: 
\begin{equation}
\label{eq: number}
N(e_1, r) = |\{e_2 \in E: dist(e_1, e_2) < r\}|
\end{equation}
\begin{equation}
\label{eq: num_cate}
N_{c}(e_1, r) = |\{e_2 \in E: dist(e_1, e_2) < r \cap e_2 \in c\}|
\end{equation}
where $e_1 \in E$, $c \in C$, and $r$ is the neighborhood radius. 
$dist(e_1, e_2)$ denotes the geographic distance between event $e_1$ 
and event $e_2$. We choose radius $r$  to be 100 meters 
as~\cite{jensen2009analyzing} did, which yields the best results in our 
final prediction performance.  
Then we can define the attractiveness value between group categories as:
\begin{equation}
\label{eq: attr}
Attr(C_a, C_b) = \frac{N - N_{C_a}}{N_{C_a}N_{C_b}} \sum_{e \in N_{C_a}} \frac{N_{C_a}(e, r)}{N(e, r) - N_{C_b}(e, r)}
\end{equation}
where $C_a$, $C_b$ are two group categories, $N$ is the total number of 
events, $N_{C_a}$ and $N_{C_b}$ are the total number of events in category 
$C_a$ and $C_b$ respectively. Here $Attr(C_a, C_b)$ represents the level 
that category $C_a$ attracts category $C_b$. Please note that 
$Attr(C_a, C_b) \neq Attr(C_b, C_a)$. Based on the definition, the qualitative assessment is: If $Attr(C_a, C_b)$ is greater than 1, events in $C_a$ have a positive attraction to events in $C_b$. Conversely, it represents a negative attractive tendency. 

Based on Jensen's attractiveness value between categories, now we can define the quality of location for event $e$ as:
\begin{multline}
\label{eq: quality}
\hat{S1}_{spatial}(e)  = \sum_{c \in \{C - C_e\}} \log(Attr(c, C_e)) \\
				\times (N_c(e, r) - \overline{N_c(e, r)})
\end{multline}
where $C_e$ is the category of event $e$, and $\overline{N_c(e, r)}$ denotes the average number of events in category $c$ that are within distance $r$ 
from the events in category $C_e$.

\subsubsection{Location Competitiveness}

Locations with higher population density may also imply more intensive competition. 
It is frequently observed that many groups with similar topics choose to 
meet in the same area, and as such events compete with each other 
to attract a shared pool of users. 
Based on this observation, we define location competitiveness in EBSN 
event organization based on the number of users (in group category 
$C_e$) whose home locations 
are within distance $R$ from a given event $e$: 
\begin{equation}
\label{eq: compete}
\hat{S2}_{spatial}(e)  = -\frac{N_{C_e}(e, R)}{N(e, R)}
\end{equation}

\subsection{Group Features}

Some recent research works have studied urban social diversity in 
location-based social networks~\cite{hristova2016measuring,noulas2015topological,cho2011friendship}. It has been observed that 
the diversity of check-ins in places to some extent reflects their popularity. Meetup groups also bring together diverse users via offline events. 
We propose two different measures to capture the diversity of group 
diversity: entropy and loyalty. 

\subsubsection{Group Member Entropy}

We employ entropy to measure the diversity of group members' interests. 
Given a group $g$, its member diversity is based on the probability of a single user $u$ attending its offline events:
\begin{equation}
p_u = \frac{|\bigcup\limits_{e \in E_g}U_e|}{\sum\limits_{e \in E_g}|U_e|}
\end{equation}
and group member entropy is defined as:
\begin{equation}
\hat{S1}_{group}(e)  = -\sum_{u \in U_g}p_u \log p_u
\end{equation}

\subsubsection{Group Member Loyalty}

Another metric for diversity of a group is whether the group's 
members have concentrated interest on the group topic, i.e., 
to what extent are the users focused on attending events 
within the same category. For each user $u$ in group $g$, we 
compute the frequency of attended events in the same 
category as the user's loyalty:
\begin{equation}
loyalty(u, g) = \frac{\sum\limits_{e \in E_u}|\{C_e = C_g\}|}{|E_u|}
\end{equation}
Then the group loyalty is measured as the average user loyalty of all active group members:
\begin{equation}
\hat{S2}_{group}(e) = \frac{\sum\limits_{u \in U_g}loyalty(u, g)}{|U_g|}
\end{equation}

\subsection{Temporal Features}
\label{sec: temporal}
Event start time is another important factor that may impact event 
popularity. For instance, some users 
may prefer to attend events after work while others only have free time 
during weekends. 


To model how well event start time matches group members' temporal preferences, we represent each event's start time as a $24 \times 7$ dimensional vector $\vec{e_t}$. Then we compute the temporal preference of each user $u \in U$ based on his/her historical event 
attendance with time decay as follows:
\begin{equation}
\label{eq: decay}
\vec{u_t} = \frac{1}{|E_u|} \sum_{e \in E_u}\frac{1}{(1+\eta)^{\theta(e)}}\vec{e_t}
\end{equation}
where $E_u$ denotes the set of historical events that user $u$ has 
participated in, $\eta$ is the time decay parameter and $\theta(e)$ denotes the number of past days. The use of the time decay function is needed 
because users' temporal preferences may change during the two-year 
period of our datasets, and more recent data would better reflect users' temporal behavior. 

Then we measure the overall satisfaction for event $e$ by adding up the Jaccard similarity between $\vec{e_t}$ and all active group members $\vec{u_t}$:
\begin{equation}
\hat{S1}_{temporal}(e)  = \sum_{u \in E_u}Jaccard(\vec{e_t}, \vec{u_t})
\end{equation}
\begin{equation}
Jaccard(\vec{e_t}, \vec{u_t}) = \frac{|\vec{e_t} \cap \vec{u_t}|}{|\vec{e_t} \cup \vec{u_t}|}
\end{equation}

\subsection{Semantic Features} 
We also propose the use of several natural language features
 to model the semantic quality of different Meetup events.

{\bf Sentiment Analysis.}
To capture the sentiment of event content, we implemented Vader~\cite{hutto2014vader},  
a lexicon and rule-based sentiment analysis tool. For each event content, it assigns a negative, neutral, or positive score based on sentiment expression.

{\bf Part-of-Speech Features.}
Given a word in event title, we can map it to its part-of-speech (POS) tag. In this paper, we propose a binary feature to measure the presence of each POS tag. The 
features we used are: adjective, adposition, adverb, conjunction, determiner, noun, numeral, particle, pronoun, verb and punctuation marks.

{\bf Text Novelty.}
We use the Jaccard similarity to identify the novelty of event titles 
by comparing it with previous event titles. 

\section{Group-based Social Influence}
\label{sec:influence}

Besides the contextual features of an event, the social influences of 
people who have RSVPed already can also affect other users' decisions 
to attend the event (thus event popularity)~\cite{goyal2010learning}. 
To utilize such 
group-specific information in EBSNs, we propose a new social 
propagation network to model people's social influences on 
event popularity that are specific to the event's group organizers.


For each event $e$, consider a directed and weighted social graph, with 
each vertex representing a Meetup user, and there exists an edge from user 
$v$ to user $u$ if $v$ RSVPed for event $e$ before $u$ did. 
The intuition is that user $v$'s RSVP for event $e$ may have  
affected user $u$'s decision to attend the same event. Furthermore, 
the influence would wane as time goes by, so the longer the time 
duration between $v$'s RSVP and $u$'s RSVP, the smaller the influence 
of $v$ on $u$. Let $N(u, e)$ be the set of users who RSVPed to $e$ 
before $u$ did, for each user $v \in N(u, e)$, we define $v$'s direct 
influence credit on $u$ as follows: 
\begin{multline}
\label{eq:decay2}
w_{v, u}(e) = \sum_{e\textprime} \frac{infl(u)}{|N(u, e\textprime)|} [\delta(G(e) = G(e\textprime)) \cdot \lambda_g \cdot {decay}_{v, u}(e\textprime) \\
+ \delta(G(e) \neq G(e\textprime)) \cdot \lambda\textprime_g \cdot {decay}_{v, u}(e\textprime)]
\end{multline} 
where $e\textprime$ denotes any event in which $v$ RSVPed before $u$. $infl(u)$ represents the fraction of activities that $u$ attended under the influence of at least one other user~\cite{goyal2010learning}. And ${decay}_{v, u}(e\textprime)$ represents the influence decays over time in an exponential tendency as:
\begin{equation}
{decay}_{v, u}(e\textprime) = exp(-\frac{t(u, e\textprime) - t(v, e\textprime)}{\tau_{v, u}})
\end{equation}
where $t(u, e\textprime)$ is the time that user $u$ RSVPed for event $e\textprime$. $\tau_{v, u}$ is the average time taken to propagate from user $v$ to user $u$. The influence decay tendency is weighted differently by 
$\lambda_g$ and $\lambda\textprime_g$, depending on whether $v$ 
and $u$ co-attended an event that was organized by the same group 
as $e$ or not.  

Using the social propagation graph, we can compute the total influence 
of user $v$ on user $u$ for event $e$: 
\begin{equation}
\Omega_{v, u}(e) = \sum_{z \in N(u, e)}\Omega_{v, z}(e)w_{z, u}(e)
\end{equation}
And the total influence that user $v$ has on all group members can be 
computed as:
\begin{equation}
\hat{S}_{influence}(v) = \sum_{u \in \{U_g - v\}}\Omega_{v, u}(e)
\end{equation}


\section{Evaluations}
\label{sec:eval} 

In this section, we evaluate the effectiveness of our proposed framework for 
predicting event popularity. 

\subsection{Methodology and Metrics}

As stated in the problem formulation, our goal is to predict the normalized popularity
value $P_e$ for each event as the overall popularity level in its group category.
Given the Meetup dataset collected in each of the three cities, we split the 
dataset into three parts.  In every city, first 80\% offline events 
of each group as the training dataset, 10\% are used for validation and 
parameter tuning, and the remaining 10\% are used for testing. 
In our CASINO framework, to integrate all context features that we have constructed, 
we fit them into Classification and Regression Tree (CART) 
model~\cite{loh2011classification}. Then we fit the residual popularity 
defined below to our social influence model: $y_e = P_e - \hat{P_e}$
The parameters in Equation~\ref{eq:decay2} are optimized by minimizing 
the least squares function $||y_e - \hat{y_e}||_2^2$ using the BFGS 
algorithm. 

We use coefficient of determination ($R^2$) as the 
evaluation metric. It is defined as:
\begin{equation}
R^2(P, \hat{P}) = 1 - \frac{\sum\limits_e(P_e - \hat{P_e})^2}{\sum\limits_e(P_e - \overline{P})^2}
\end{equation}
where $\overline{P}$ is the mean of $P$. 
For the testing procedure, the final results we report are computed by: $R^2(P_e, \hat{P_e} + \hat{y_e})$. 

We compare our CASINO framework with the following approaches: 
(1) \textbf{NM} is a naive-mean based method that predicts future 
event popularity $\hat{P_e}$ as the average of historical event 
popularity of the same group; 
(2) \textbf{SVD-MFN~\cite{du2014predicting}} is a state-of-the-art 
context-aware event attendance prediction algorithm for individual users 
and we use its predictions for individual users to compute the overall 
popularity of each event;  
(3) \textbf{Cont} uses only our contextual features to predict 
$P_e$ directly; and 
(4) \textbf{CASINO(-)} uses both contextual features and 
group-based social influence without considering group difference 
(i.e., $\lambda_g = \lambda_g\textprime$). 

\subsection{Overall Prediction Performance}

Table~\ref{tab: evaluation} summarizes the event popularity prediction performance of different approaches using the $R^2$ metric in three cities. 
The baseline approach NM is ineffective and only has an average $R^2$ 
of 0.165 across three cities. SVD-MFN did not achieve good results either, 
one possible reason is 
that event participation is highly skewed and most users do not participate 
in a given event. In contrast, our combined framework can provide much better prediction results. Our CASINO framework performances  best in all three cities, achieving 0.758 for New York, 0.723 for London and 0.718 for Sydney. It improves the prediction performance by 130\% over the baseline approach. In addition, the improvement from 
Cont to CASINO(-) and to CASINO demonstrate the effectiveness 
of our contextual features, the social influence feature, and the importance 
of differentiating social influences for different groups. 


\begin{table}[]
		\centering
		
		\small
\caption{Performance Comparison of Different Models for Event Popularity Prediction}		
		\begin{tabular}{|c|c|c|c|c|c|}
			\hline
			    & NM                   & SVD-MFN                     & Cont                 & CASINO(-)                & CASINO                 \\ \hline \hline 
			NYC          & 0.240          & 0.319           & 0.730          & 0.744           & \textbf{0.758}           \\ \hline
			LON          & 0.140          & 0.305           & 0.672          & 0.692           & \textbf{0.723}           \\ \hline
			SYD          & 0.117          & 0.289       & 0.653           & 0.685         & \textbf{0.718}           \\ \hline
						
		\end{tabular}
\label{tab: evaluation}

\end{table}

As discussed in the spatial and temporal features subsections, there are two 
parameters in our context model: radius $R$ and time decay $\eta$. They are determined by a grid search on our validation set. The specific parameters values for New York, London and Sydney are the 
following:  radius $R$ is set to 1.5 miles and the time decay parameter $\eta$ is set to 0.01 for all three cities. 

\section{Conclusions}
\label{sec:conclusion}

In this work, we have studied the problem of event popularity in 
EBSNs and developed four contextual models 
(spatial, group, temporal, and semantic) and a group-based 
social influence model for analyzing and predicting the popularity 
of events organized by different social groups. 
Our combined CASINO framework achieves high prediction accuracy for 
real-world Meetup datasets collected in three major cities around the world. 
We further analyze the contributions of individual models and the impacts 
of different event organization scenarios. Our study offers initial 
new insights for event organizers as well as targeted advertising strategies
for EBSN service providers. 

\section{Acknowledgements}
This work is supported in part by the US National Science Foundation (NSF) through grant CNS 1528138.

\bibliographystyle{aaai}
\bibliography{citation}

\begin{thebibliography}{}

\bibitem[\protect\citeauthoryear{Cho, Myers, and
  Leskovec}{2011}]{cho2011friendship}
Cho, E.; Myers, S.~A.; and Leskovec, J.
\newblock 2011.
\newblock Friendship and mobility: user movement in location-based social
  networks.
\newblock In {\em KDD},  1082--1090.

\bibitem[\protect\citeauthoryear{Du \bgroup et al\mbox.\egroup
  }{2014}]{du2014predicting}
Du, R.; Yu, Z.; Mei, T.; Wang, Z.; Wang, Z.; and Guo, B.
\newblock 2014.
\newblock Predicting activity attendance in event-based social networks:
  Content, context and social influence.
\newblock In {\em UbiComp},  425--434.

\bibitem[\protect\citeauthoryear{Georgiev, Noulas, and
  Mascolo}{2014a}]{ICWSM148071}
Georgiev, P.; Noulas, A.; and Mascolo, C.
\newblock 2014a.
\newblock Where businesses thrive: Predicting the impact of the olympic games
  on local retailers through location-based services data.
\newblock In {\em ICWSM},  151--160.

\bibitem[\protect\citeauthoryear{Georgiev, Noulas, and
  Mascolo}{2014b}]{georgiev2014call}
Georgiev, P.~I.; Noulas, A.; and Mascolo, C.
\newblock 2014b.
\newblock The call of the crowd: Event participation in location-based social
  services.
\newblock In {\em ICWSM},  141--150.

\bibitem[\protect\citeauthoryear{Goyal, Bonchi, and
  Lakshmanan}{2010}]{goyal2010learning}
Goyal, A.; Bonchi, F.; and Lakshmanan, L.~V.
\newblock 2010.
\newblock Learning influence probabilities in social networks.
\newblock In {\em WSDM},  241--250.

\bibitem[\protect\citeauthoryear{Hristova \bgroup et al\mbox.\egroup
  }{2016}]{hristova2016measuring}
Hristova, D.; Williams, M.~J.; Musolesi, M.; Panzarasa, P.; and Mascolo, C.
\newblock 2016.
\newblock Measuring urban social diversity using interconnected geo-social
  networks.
\newblock In {\em WWW},  21--30.

\bibitem[\protect\citeauthoryear{Hutto and Gilbert}{2014}]{hutto2014vader}
Hutto, C.~J., and Gilbert, E.
\newblock 2014.
\newblock Vader: A parsimonious rule-based model for sentiment analysis of
  social media text.
\newblock In {\em ICWSM},  216--225.

\bibitem[\protect\citeauthoryear{Jensen}{2009}]{jensen2009analyzing}
Jensen, P.
\newblock 2009.
\newblock Analyzing the localization of retail stores with complex systems
  tools.
\newblock In {\em Proc. of the 8th Intl. Symposium on Intelligent Data
  Analysis},  10--20.

\bibitem[\protect\citeauthoryear{Loh}{2011}]{loh2011classification}
Loh, W.-Y.
\newblock 2011.
\newblock {Classification and Regression Trees}.
\newblock {\em Wiley Interdisciplinary Reviews: Data Mining and Knowledge
  Discovery} 1(1):14--23.

\bibitem[\protect\citeauthoryear{Macedo, Marinho, and
  Santos}{2015}]{macedo2015context}
Macedo, A.~Q.; Marinho, L.~B.; and Santos, R.~L.
\newblock 2015.
\newblock Context-aware event recommendation in event-based social networks.
\newblock In {\em RecSys},  123--130.

\bibitem[\protect\citeauthoryear{Noulas \bgroup et al\mbox.\egroup
  }{2015}]{noulas2015topological}
Noulas, A.; Shaw, B.; Lambiotte, R.; and Mascolo, C.
\newblock 2015.
\newblock Topological properties and temporal dynamics of place networks in
  urban environments.
\newblock In {\em WWW},  431--441.

\end{thebibliography}

\end{document}